# Large magnetodielectric response in $Pr_{0.6}Ca_{0.4}MnO_3$/ polyvinylidene fluoride nanocomposites

K. Devi Chandrasekhar[1], A.K.Das[1] and A.Venimadhav[2,a]

[1]*Department of Physics & Meteorology, Indian Institute of Technology, Kharagpur-721302*
[2] *Cryogenic Engineering Centre, Indiana Institute of Technology, Kharagpur-721302*

## Abstract

We have studied the magnetic field effect on low frequency dielectric properties of $Pr_{0.6}Ca_{0.4}MnO_3$/polyvinylidene fluoride nanocomposite with 22.5% volume fraction of $Pr_{0.6}Ca_{0.4}MnO_3$ nanoparticles. A strong magnetodielectric response was observed below 120 K where $Pr_{0.6}Ca_{0.4}MnO_3$ nanoparticles show the magnetic phase transition indicating a direct correlation between magnetism and dielectric properties. A large change of the dielectric permittivity ~ 30% has been observed in a magnetic field of 4.6 T with loss as low as 0.17 at 70 K. The observed magnetodielectric response has been attributed to the decrement of polaron activation barrier of $Pr_{0.6}Ca_{0.4}MnO_3$ nanoparticles with the increase of magnetic field.

[a] Corresponding author: venimadhav@hijli.iitkgp.ernet.in

Recently multiferroics have received enormous research attention as they can bring magnetic control of electrical polarisation and electric control of magnetization as an extra degree of freedom in device utilization.[1] The simultaneous presence of two or more ferroic properties in a single phase material has been found to be challenging.[1] The composite systems offer an interfacial strain induced multiferroic effect having large coupling coefficients in comparison with the single phase systems.[1] One of the ways to quantify multiferroic behaviour is the observation of dielectric polarisation anomaly at the magnetic transition temperature (magnetodielectric effect). Though such dielectric anomaly has been observed in various oxides like $BiMnO_3$,[2] $SeCuO_3$ and $TeCuO_3$, the magnitude of this effect was very small.[3] On the other hand, magnetodielectric effect has also been proposed and shown to occur in systems with the combination of magnetoresistance and Maxwell-Wagner interfacial polarisation.[4] Recently, this extrinsic effect induced magnetodielectric behaviour has been observed in polycrystalline samples and phase separated manganites.[4,5] Besides a large magnetodielectric effect, unfavourable loss also increases due to the magnetoresistance that restricts the applicability of these polycrystals for practical devices.

Multiphase heterogeneous system is a good choice for engineering the magnetodielectric effect with minimal loss. In this letter, we present the dielectric and magnetodielectric response of $Pr_{0.6}Ca_{0.4}MnO_3$ (PCMO)/polyvinylidene fluoride (PVDF) composite with 22.5% volume fraction of PCMO nanoparticles. A large magnetodielectric effect of 30% at 100 kHz with a minimum loss of ~ 0.17 was obtained at 70 K with the applied magnetic field of 4.6 T. PCMO is a known charge ordered manganite that shows colossal magnetoresistance (CMR) and colossal electroresistance (CER) effect because of strong spin, charge and orbital coupling.[6,7] Recently Ch. Jooss et al., observed the appearance of local ferroelectric domain through TEM analysis on $Pr_{1-x}Ca_xMnO_3$ ($0.32 \leq x \leq 0.5$) bulk samples and films and assigned the ferroelectric polarization to Zener polaron order states.[8] The insulating PVDF polymer plays a key role to achieve the mechanical flexibility and reducing the undesired loss of CMR PCMO filler by an order of magnitude.

PCMO nanoparticles were synthesized using sol-gel based polymeric precursor route.[9] PVDF/PCMO nanoblends were prepared by mixing desired amounts of PCMO nanoparticles and PVDF in methanol and fabricated pellets which were annealed at 200°C mainly to reduce porosity. Such nanocomposites showed a large dielectric permittivity at the percolation point due to enhanced interfacial polarization effect.[9] Dielectric properties were done using HIOKI LCR Bridge 3532-50 with an excitation ac voltage of 3V by applying the silver paste contacts on both sides of the sample to make parallel plate capacitor geometry. The temperature and the magnetic field variation of dielectric properties as a function of frequency were performed using a closed cycle cryogen free superconducting system from 300 K to 5 K. The disk type sample was placed parallel to the direction of magnetic field. The temperature dependent dc magnetic measurements were performed on PCMO nanoparticles using Quantum Design DC VSM SQUID magnetometer.

Fig. 1(a) shows the zero field cooling (ZFC) and field cooled (FC) magnetization as a function of temperature performed with a magnetic field of 100 Oe. The ZFC and FC magnetization curves exhibit strong irreversibility with a peak in the ZFC curve defined as $T_B$. The FC magnetization increases strongly below the irreversibility temperature $T_{irr}$, where the ZFC and FC curves merge. The magnetization data exhibits two distinct magnetic transitions around ~ 114 K and ~ 42.9 K ($T_f$). The bulk PCMO exhibits the charge ordering transition (CO) at 240 K and an antiferromagnetic transition at 170 K. However, these two properties get suppressed in the case of nanopartilces.[10] The large bifurcation, appearance of cusp in ZFC data and continuous increase of FC magnetization are common features of some of the metastable magnetic systems such as cluster glass and superparamagnetic particles. The magnetic behaviour of the PCMO nanoparticles is similar to that of nano $Fe_3O_4$ particles and nanoparticles of CMR manganites, where the magnetic behaviour was understood using a core-shell model.[11] The details about the nature of magnetic transitions and the associated glassy behaviour will be reported elsewhere.

Temperature dependent real part of dielectric permittivity and the loss tangent as a function of frequency are shown in the Fig. 1(b). As shown in the figure, the dielectric permittivity decreases with the decrease of temperature and exhibits step like features, one at ~ 210 K and another at ~ 80 K. Below 20 K, the dielectric permittivity exhibits a saturating value ~ 5 irrespective of the frequency and this can be regarded as the intrinsic dielectric permittivity of the PCMO/PVDF composites. At higher temperatures, the real part of the dielectric permittivity shows frequency dependent dispersion. The step like features in dielectric permittivity are accompanied by broad relaxation peaks in the loss tangent spectrum and these peak positions are shifted towards higher temperature with the increase of frequency. The high temperature relaxation process in the paramagnetic (PM) region (~ 170 K) can be ascribed to the insulating barriers separated semiconducting grain boundary response.[12] Whereas the low temperature dielectric relaxation in the ferromagnetic (FM) region can be related to the hopping of polaron charge carriers at the localized sites and such a relaxation behaviour has been observed in many manganite systems.[9, 13, 14] The glass transition relaxation temperature of PVDF that occurs at 240 K has not been observed; this indicates that the dominant dielectric response is coming from that of PCMO nanoparticles. These relaxation phenomena have been fitted to the Arrhenius law given by the equation (1)

$$\tau = \tau_o \exp\left(\frac{E_a}{k_B T}\right) \quad (1)$$

where $\tau_o$ is the pre-exponential factor, $E_a$ is activation energy, and $K_B$ is Boltzmann constant. A fit to equation (1) in the paramagnetic region (160-300 K) yields the activation energy of 234 ± 3.14 meV with $\tau_o$ = 2.085 x $10^{-10}$ (not shown in the figure). However, the polaron hopping activation energy in the ferromagnetic region 40 K ≤ T ≤ 80 K is shown in the inset of figure 2(a) and the fitting parameters $\tau_o$ and $E_a$ are obtained as 4.92 x 10$^{-11}$ sec and ~ 80.3 ± 2.27 meV respectively. The obtained activation energies in the paramagnetic and ferromagnetic regions are comparable to that of polaron activation energies of the bulk PCMO.[7, 13]

To study the effect of applied magnetic field on the dielectric properties, we have performed temperature dependent dielectric measurements as a function of magnetic field. As shown in the Fig. 2 the real part of the effective dielectric permittivity and loss tangent shows similar kind of temperature dependence as that of the zero field dielectric measurement (ZFD). However, application of magnetic field shows divergence of both dielectric permittivity and the loss tangent below ~ 120 K and interestingly, this match with the magnetic transition of the PCMO nanoparticles. This indicates a strong correlation between magnetism and the dielectric properties. Moreover, the step like feature in the dielectric permittivity and the corresponding loss peak are shifted systematically to the low temperature side with the increase of magnetic field. The magnetodielectric (MD) response was absent below ~ 20 K. In order to understand the origin of this MD behaviour, the polaron induced activation energies for different applied fields has been obtained by fitting the relaxation data to equation (1) as shown in the inset of the Fig. 2a. The parameters $\tau_o$ and $E_a$ corresponding to different magnetic fields obtained from the fit are plotted against the applied magnetic field is shown in the inset of Fig. 2b. As shown in the figure, the activation energy decreases slowly with the increase of the magnetic field up to 3 T and above that a large drop in activation energy is observed. Similarly, $\tau_o$ is independent of applied magnetic field up to 3 T followed by a sudden raise at higher magnetic fields. The decrease of activation energy with the application of external magnetic field may be assigned to the magnetic field assisted polaron melting. Recently, Ch. Jooss *et al.* performed the HRTEM analysis on PCMO polycrystals, and confirmed that the melting of polarons is the key factor for observing the colossal electroresistance and colossal magnetoresistance behaviour in PCMO manganites.[8] In the PCMO/PVDF nanocomposites, application of external magnetic field melts the polarons, and this leads to a large change in the dielectric properties. Here high insulating PVDF plays key role to reduce the high loss induced by CMR behaviour of the PCMO fillers. However, the activation energy (229.6 ± 6.45 meV) in the PM region is unaffected by the external magnetic fields up to 4.6T (not shown here).

The magnetic field dependence of MD response at 100 kHz for different temperatures is shown in the Fig. 3, where the MD effect is defined as

$$\text{MD (\%)} = \left[\frac{\varepsilon_{4.6T} - \varepsilon_{0T}}{\varepsilon_{0T}}\right] * 100 \qquad (2)$$

As shown in the Fig. 3, the MD (%) increases with the increase of magnetic fields up to 4.6 T. We have also found that the dielectric permittivity did not retrace in its original path while sweeping back the field from 4.6 T to zero. And this hysteresis behaviour has been found to be independent of applied frequencies. However, the width of the loop decreases with increase of temperatures and vanished completely at 110 K. The MD (%) increase with the increase of temperature reaches a maximum value of 30% at 70 K in a magnetic field of 4.6 T and then decreases. The origin of the observed hysteresis behaviour has been assigned to the kinetic arrest of ferromagnetic metallic phase in the antiferromagnetic insulating phase.[10] The intrinsic magnetodielectric coupling in a magnetoelectric medium has been understood based on the Ginzburg-Landau phenomenological theory. According to this theory, below $T_c$ the magnetodielectric effect is proportional to the spin-pair correlation of neighboring spins, hence the square of the magnetization MD (%) or $\Delta\varepsilon = \gamma M^2$. In fact, near PM to FM transition (~ 110 K) we do see the validity of the above relation like in $BiMnO_3$ (plot not shown). Interestingly, in the FM region where the maximum MD effect is observed, the MD (%) effect follows a linear relationship with $M^4$. This kind of higher order magnetoelectric electric coupling has been observed in systems with reduced dimensionality such as multilayer thin films. [15] However, $Pr_{0.6}Ca_{0.4}MO_3$ can be realized as a relaxor composed of nano scale ordered Zener polaron and disordered Jahn Teller polaron phases where the ordered phase having electric polarization due to non-centrosymmetric displacements of the $MnO_6$ octahedral. Below the magnetic ordering (in the FM region) a higher order coupling between electric dipoles and magnetization can be possible via magnetic double exchange interaction. More experiment and theoretical understanding is necessary to uncover the reason behind higher order coupling in this system.

In Summary, we have investigated the magnetic field effect on the dielectric properties of the PCMO/PVDF nanocomposite with 22.5% volume fraction of PCMO nanoparticles. A large change of the effective dielectric permittivity of ~ 30% has been observed at 70 K in a magnetic field of 4.6 T with loss as low as 0.17. The observed MD effect has been attributed to the decrease of polaron activation barrier under magnetic field. The high magnetodielectric effect with minimal loss and mechanical flexibility in PCMO/PVDF shows that the manganite-polymer nanocomposites are potential for multifunctional applications.

This work was supported by CSIR grant-22(0465)/09/EMR-II and the authors also acknowledge IIT Kharagpur for funding VSM SQUID magnetometer and DST, New Delhi for FIST grant for establishing Cryogen free high magnetic field facility.

Figure captions:

FIG. 1: (a) Temperature variation of ZFC-FC magnetization of PCMO nanoparticles. (b) Temperature variation of effective dielectric permittivity of PCMO/PVDF nanocomposite; inset shows the temperature dependence of loss tangent.

Fig. 2: (a) Temperature variation of effective dielectric permittivity of the PCMO/PVDF nanocomposite for different magnetic fields; inset show the plot between the inverse temperature and the relaxation time for different applied fields; solid line is fitting curve for the equation (1) (b) Temperature variation of loss tangent for different applied magnetic fields. Inset shows the variation of $E_a$ and the $\tau_o$ with respect to the magnetic field.

Fig. 3: (a) Magnetic field variation of MD (%) at 100 kHz; inset shows the plot between the $M^4$ and MD (%) vs. H.

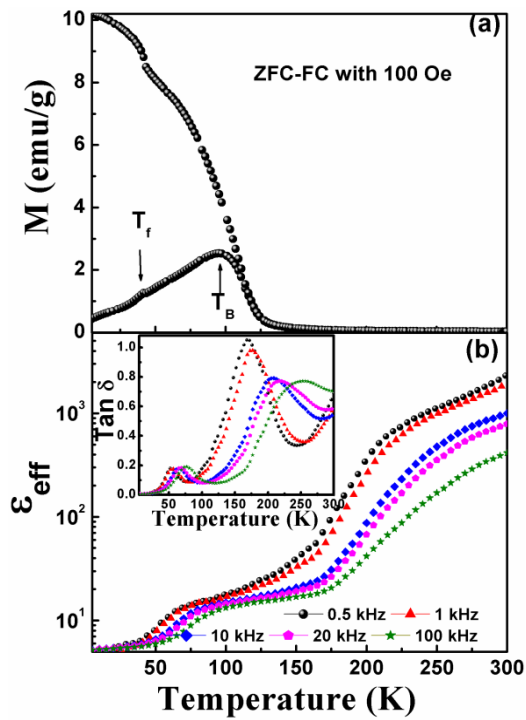

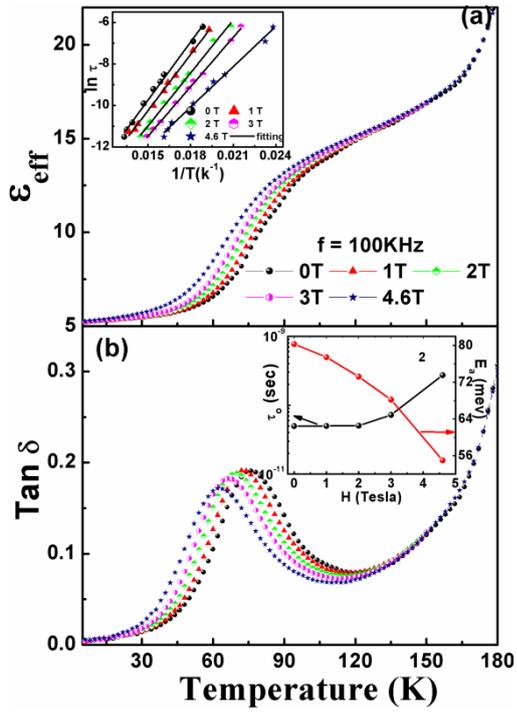

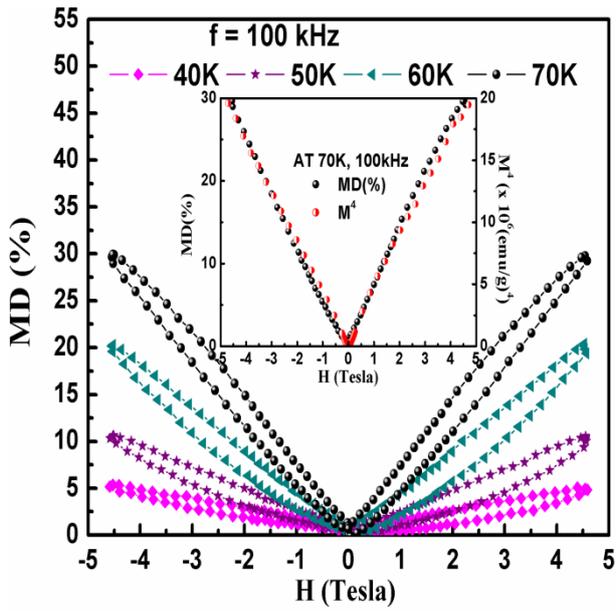